\documentclass[12pt]{nature}

\usepackage{scicite}

\usepackage{times}
\usepackage{amsmath}
\usepackage{graphicx}

\usepackage{txfonts}
\usepackage{color}
\usepackage[version=4]{mhchem}

\newenvironment{sciabstract}{%
\begin{quote} \bf}
{\end{quote}}

\title{Chiral emission of vortex microlasers enabled by collective modes 
of guided resonances}

\author
{Ye Chen$^{1,\#}$, Mingjin Wang$^{2,4,5,\#}$, Jiahao Si$^{2}$, Zixuan Zhang$^{1}$, Xuefan Yin$^{1}$, Jingxuan Chen$^{2}$, NianYuan Lv$^{1}$,  Chenyan Tang$^{2}$, Wanhua Zheng$^{2,4,5,6,7,\ast}$, Yuri Kivshar$^{3,\ast}$, Chao Peng$^{1,8\ast}$\\
\\
\normalsize{$^{1}$State Key Laboratory of Advanced Optical Communication Systems and Networks, School of Electronics \& Frontiers Science Center for Nano-optoelectronics, Peking University, Beijing 100871, China}\\
\normalsize{$^{2}$Laboratory of Solid State Optoelectronics Information Technology, Institute of Semiconductors, CAS, Beijing 100083, China}\\
\normalsize{$^{3}$Nonlinear Physics Centre, Research School of Physics,
The Australian National University, Canberra ACT 2601, Australia}\\
\normalsize{$^{4}$State Key Laboratory on Integrated Optoelectronics, Institute of Semiconductors, CAS, Beijing 100083, China }\\
\normalsize{$^{5}$Hangzhou Institute for Advanced Study, University of Chinese Academy of Sciences, Hangzhou 310024, China }\\
\normalsize{$^{6}$College of Future Technology, University of Chinese Academy of Sciences, Beijing 101408, China}\\
\normalsize{$^{7}$Center of Materials Science and Optoelectronics Engineering, University of Chinese Academy of Sciences, Beijing 100049, China}\\
\normalsize{$^{8}$Peng Cheng Laboratory, Shenzhen 518055, China}\\
\\
\normalsize{$^\#$These authors contributed equally to this work.}\\
\normalsize{$^\ast$To whom correspondence should be addressed; E-mail:    Wanhua Zheng (whzheng@semi.ac.cn), Yuri Kivshar (yuri.kivshar@anu.edu.au), Chao Peng (pengchao@pku.edu.cn). }\\\\
}

\date{}

\begin{document} 


\baselineskip24pt


\maketitle

\begin{sciabstract}
Vortex lasers have attracted substantial attention in recent years owing to their wide array of applications such as micromanipulation, optical multiplexing, and quantum cryptography. In this work, we propose and demonstrate chiral emission of vortex microlaser leveraging the collective modes from omnidirectionally hybridizing the guided mode resonances (GMRs) within photonic crystal (PhC) slabs. Specifically, we encircle a central uniform PhC with a heterogeneous PhC that features a circular lateral boundary. Consequently, the bulk GMRs hybridize into a series of collective modes due to boundary scatterings, resulting in a vortex pattern in real space with a spiral phase front in its radiation. Benefiting from the long lifetime of GMRs as quasi-bound state in the continuum and using asymmetric pumping to lift the chiral symmetry, we demonstrate stable single-mode lasing oscillation with a low optical pumping threshold of $18~\mathrm{kW/cm^2}$ at room temperature. We identify the real-space vortex through polarization-resolved imaging and self-interference patterns, showing a vivid example of applying collective modes to realize compact and energy-efficient vortex microlasers.

\end{sciabstract}

% In setting up this template for *Science* papers, we've used both
% the \section* command and the \paragraph* command for topical
% divisions.  Which you use will of course depend on the type of paper
% you're writing.  Review Articles tend to have displayed headings, for
% which \section* is more appropriate; Research Articles, when they have
% formal topical divisions at all, tend to signal them with bold text
% that runs into the paragraph, for which \paragraph* is the right
% choice.  Either way, use the asterisk (*) modifier, as shown, to

\section*{Introduction}

An optical vortex is a beam of photons that travels with a helical phase wavefront and carries quantized orbital angular momentum (OAM)\cite{allenOrbitalAngularMomentum1992}. Due to their unique ability to spatially differentiate photons, optical vortices find applications in optical sensing\cite{laveryDetectionSpinningObject2013}, micromanipulation\cite{patersonControlledRotationOptically2001,macdonaldCreationManipulationThreeDimensional2002}, and optical communications in the classical\cite{wangTerabitFreespaceData2012,leiMassiveIndividualOrbital2015,xieUltrabroadbandOnchipTwisted2018} and quantum\cite{chenIntegratedOpticalVortex2024} realms. In recent years, many efforts have paid to shrinking vortex generators to a compact footprint to facilitate chip-scale data transmission and quantum processing\cite{caiIntegratedCompactOptical2012,liuDirectFiberVector2017,shaoSpinorbitInteractionLight2018,xieUltrabroadbandOnchipTwisted2018,zhouUltracompactBroadbandPolarization2019,sroorHighpurityOrbitalAngular2020,wangVortexRadiationSingle2022,meiCascadedMetasurfacesHighpurity2023,chenIntegratedOpticalVortex2024,liuIntegratedVortexSoliton2024}. Exciting advances in vortex microlasers have emerged using phase singularities in real\cite{liOrbitalAngularMomentum2015,miaoOrbitalAngularMomentum2016,carlonzambonOpticallyControllingEmission2019,sunLeadHalidePerovskite2020,yangSpinMomentumLockedEdgeMode2020,zhangTunableTopologicalCharge2020,zhangUltrafastControlFractional2020,chenBrightSolidstateSources2021,piccardoVortexLaserArrays2022,zhangSpinOrbitMicrolaser2022} or momentum\cite{huangUltrafastControlVortex2020,mohamedControllingTopologyPolarization2022,hwangVortexNanolaserBased2023} spaces. For example, in real-space implementations, microring resonators incorporated angular grating structures\cite{miaoOrbitalAngularMomentum2016,zhangTunableTopologicalCharge2020,zhangUltrafastControlFractional2020,chenBrightSolidstateSources2021, zhangSpinOrbitMicrolaser2022} and vertical-cavity surface-emitting devices\cite{liOrbitalAngularMomentum2015, carlonzambonOpticallyControllingEmission2019,sunLeadHalidePerovskite2020} using planar spiral nanostructures extract the OAM modes into free space. Although real-space vortex microlasers are compact with notable emission directionality and great efficiency, they face challenges from significant scattering losses that reduce the quality factors ($Q$s) of laser cavities. In contrast, momentum-space implementations utilize the Pancharatnam-Berry (PB) phase in PhC slabs to generate vortex beams\cite{huangUltrafastControlVortex2020}. By employing photonic bound states in the continuum (BICs)\cite{hsuObservationTrappedLight2013,zhenTopologicalNatureOptical2014,jinTopologicallyEnabledUltrahighQ2019}, namely, a class of discrete states with infinite lifetimes embedded in the radiation continuum, momentum-space vortex lasers exhibit low lasing thresholds due to the high-$Q$ nature of BICs\cite{huangUltrafastControlVortex2020}, but they are relatively bulky in size and have difficulties in supporting high-order vertices since the lattice symmetry restrains the PB phases. 

Recently, the advance of vortex generation in a photonic disclination cavity shows the remarkable light control capability of periodic photonic structures\cite{hwangVortexNanolaserBased2023}, inspiring the potential of PhC to achieve spiral phases in real space, even though the light path is not explicitly defined. It was reported that a group of individual resonances can collectively oscillate as a whole and exhibit distinctive features that are different from single resonances\cite{auguieCollectiveResonancesGold2008,kuznetsovOpticallyResonantDielectric2016,limonovFanoResonancesPhotonics2017,yesilkoyUltrasensitiveHyperspectralImaging2019,hoangCollectiveNatureHighQ2024} . For instance,  an array of nanoparticles with Mie resonances localized in real space can couple via the far field, leading to collective modes in the momentum space. According to the law of duality, we expect that the GMRs with discrete momentum can form collective modes in real space,  thereby paving the way for compact and low-threshold vortex microlasers for on-chip integration.

In this work, we propose and demonstrate chiral emission of vortex microlaser by utilizing a class of collective modes of GMRs in PhC slabs, in which the optical modes circulate in real space and radiate as helical phase fronts. Specifically, we design a square lattice PhC encircled by a circular lateral boundary on an active InGaAsP membrane. As a consequence of the circular boundary's scattering, the GMRs that originally propagated along specific Bloch wavevectors in bulk PhC omnidirectionally hybridize into two-fold degenerate collective modes with opposite chirality, and they are favorable for lasing because of the protection of quasi-BICs. By applying the technique of asymmetric pumping to break chiral symmetry,  we experimentally demonstrate stable single-mode lasing behavior with an optical-pumping threshold of  $18~\mathrm{kW/cm^2}$, and identify the phase vortex features in real space through polarization-resolved imaging and self-interference patterns.

\section*{Principle and design}
We schematically present our vortex microlaser design in Fig. 1a, in which square latticed air holes are patterned on a suspended InGaAsP membrane. Six layers of multiple quantum wells (MQWs) are embedded in the center of the membrane as optical gain materials at the C-band of the telecom wavelength. The details of the MQW epitaxial wafer are presented in the Suppl. Section 1. We use the photonic bandgap (PBG) to localize the light in the transverse direction \cite{chenObservationMiniaturizedBound2022}. Namely, the central region (region A with the radius $r_A$) is surrounded by a heterogeneous PhC (region B with the radius $r_B$) to form a circular boundary of PBG for lateral confinement. Then, the scatterings at the circular boundary would drive the bulk GMRs that are aligned with the iso-frequency contour in momentum space to collectively oscillate. As a result, we found a set of collective modes that counter-intuitively circulate in real space, enabling a helical phase wavefront upon the radiation in the out-of-plane direction.  

To elaborate on the principle of vortex generation, we first present the detailed geometry and band diagram of the designed PhCs (Fig. 1b), focusing on the transverse electric band A (TE-A) of region A in the vicinity of the 2nd-order $\Gamma$ point within the bandgap of region B. It is worth noting that, in infinite periodic PhCs, the bulk GMRs $|k\rangle$ and $|k'\rangle$ with Bloch wavevectors $k\neq k'$ are mutually orthogonal. Consequently, they propagate independently in straight paths along specified directions due to the translational symmetry of the PhC (left panel, Fig. 1c), preventing them from altering propagation directions to create a well-defined closed-loop in the real space like in microring resonators. However, the introduction of a non-uniform boundary in the PhC can give rise to additional momenta $\Delta k = k - k'$\cite{notomiTheoryLightPropagation2000,notomiNegativeRefractionPhotonic2002}, leading to the coupling between $|k\rangle$ to $|k'\rangle$ according to the scattering effect (right panel, Fig. 1c). Essentially, the modes within the circular boundary PhC can be viewed as a collectively linear combination of a set of bulk GMRs due to the boundary's scattering (known as collective modes), allowing for well-defined energy flows in the real space to generate unique phase fronts in the radiation (lower-panel, Fig. 1c). 

Specifically, we present a 3D visualization of the TE-A band's dispersion in Fig. 1d, revealing a quadratic curvature near the $\Gamma$ point. To ensure the harmonic stability of collective modes in the time domain, the GMRs involved in the linear combination must possess identical frequencies, namely they should align with the iso-frequency contour of the band. As depicted in Fig. 1d (red circle), this contour is nearly circular with a radius of $\Delta k_r=\mu_{m}^{l}/R$, where $\mu_{m}^{l}$ denotes the $l$-th zeros of the  $m$-th Bessel function $J_m(\mu_{m}^{l})=0$ and $R$ denotes the cavity's radius in real space. When $R$ is sufficiently large, the real space boundary tends towards roundness, irrespective of the discreteness of the PhC lattice. Simultaneously, the contour of $\Delta k_r$ in momentum space shrinks, resulting in nearly isotropic band dispersion.  Based on our design featuring a circular boundary of $R=10a$, we observe that both the boundary and dispersion remain almost invariant under the rotations of any angle. Consequently, the bulk GMRs ranging from $|k,0^\circ\rangle$ to $|k,360^\circ\rangle$ on the iso-frequency contour propagate and scatter omnidirectionally as illustrated schematically in Fig. 1d (see Suppl. Section I for detail discussions), which facilitates their hybridization into a set of collective modes that retains continuous rotation symmetry akin to a microring resonator, leading to paired, degenerate modes rotating clockwise (CW) and counter-clockwise (CCW) in real space without the requirements for physical light transmission paths. Furthermore, through the diffraction of the PhC lattice, a spiral phase front of radiation, namely a vortex beam, is generated.

The collective modes are identified via numerical simulation (COMSOL Multiphysics, Fig. 2a), consistent with the analytical solutions derived from the coupled-wave theory (CWT)\cite{liangThreedimensionalCoupledwaveModel2011,chenAnalyticalTheoryFinitesize2022} in polar coordinates (see the details in Suppl. Section 2). The slow-varying envelopes of collective modes are represented by the Bessel function $J_m(r)$, denoted by the quantum numbers $m$ and $l$ in the azimuth and radial directions, respectively. It's worth noting that, except for $m=0$,  each quantum number $(m,l)$ corresponds to a pair of two-fold degenerate modes that rotate in either CW or CCW directions. As an illustration, we choose the CW mode of $(1,1)$ for further analysis, which is our candidate for lasing oscillation. We calculate the snapshot magnetic field at fixed time intervals within one oscillation cycle, ranging from $0$ to $T$, clearly demonstrating the rotating in real space (Fig. 2b). This rotational motion generates a helical phase front spanning from $0$ to $2\pi$ in out-of-plane radiation, akin to the motion of a propeller driving water. Furthermore, the distribution of electrical field strength ($|E|$) is plotted in the center of Fig. 2b, showing a donut-shape pattern in real space, as a consequence of temporal averaging of energy flow. When a linear polarizer is placed before observing the donut beam, two lobes with equal density intensities would appear along the direction perpendicular to the polarizer's major axis, thus providing a distinctive feature to identify the vortex beam (Fig. 2c).

The circular collective modes have the potential to achieve low-threshold lasing because their radiation losses can be suppressed by using quasi-BICs\cite{chenObservationMiniaturizedBound2022,renLowthresholdNanolasersBased2022}. By tuning the structural parameters $a$ and $r_A$, we found eight off-$\Gamma$ BICs and one symmetry-protected BIC in the momentum space, carrying integer topological charges on their far-field polarization (Fig. 3a). We calculate the $Q$s of the TE-A band in the bulk PhC (Fig. 3b), showing that the lifetime diverges to infinity at $k_{\text{BIC}}=0.034$, giving rise to a ring of high $Q$ values at $|k_{\text{BIC}}|=0.034$.  As we reported previously\cite{chenObservationMiniaturizedBound2022, renLowthresholdNanolasersBased2022,chenAnalyticalTheoryFinitesize2022}, the radiations in finite-size PhC can be highly directional, determined by the sizes and shapes of the boundaries. With a circular boundary, the radiations show equal weights along the contour of $\Delta k_r$, and they can be effectively suppressed by aligning the in-plane momentum of radiation with the high-$Q$ ring as $\Delta k_r=k_{\text{BIC}}$ (Fig. 3a). Further, we fine-tune the structural parameters of circular the PBG in region B to best suppress the radiation while keeping the mode wavelength fixed at $\sim 1550$ nm. As presented in Fig. 3c, we obtain an optimal $Q$ of $\sim 10^6$ at the parameters of $a=537$ nm, $r_{A}=164$ nm, and $r_{B}=154$ nm, which is sufficient for low-threshold lasing.

When the PhC cavity is perfectly round, the boundary and dispersion own continuous rotational symmetry, allowing for the two-fold degeneracy of CW and CCW modes, represented by $|\sigma_\pm\rangle$, respectively, which have identical complex eigenfrequencies $\lambda_\pm$ described by $\mathbf{H} |\sigma_\pm\rangle = \lambda_\pm |\sigma_\pm\rangle$, where $\mathbf{H}$ denotes the Hamiltonian of the system.  However, in realistic samples, imperfections in fabrication inevitably affect the real part of the permittivity, while the asymmetric alignment during optical pumping introduces non-uniform gain, leading to a spontaneous chiral symmetry breaking depicted by a perturbation of Hamiltonian $\Delta \mathbf{H}$ (Fig. 3d). Accordingly, the CW and CCW modes would experience a non-Hermitian mutual coupling of $\kappa = \langle \sigma_\mp | \mathbf{\Delta H} | \sigma_\pm \rangle$ that lifts the degeneracy, giving rise to different frequencies and $Q$s of $| \sigma_\pm \rangle$\cite{wiersigAsymmetricScatteringNonorthogonal2008,wiersigNonorthogonalPairsCopropagating2011,wiersigStructureWhisperinggalleryModes2011,pengChiralModesDirectional2016,chenExceptionalPointsEnhance2017}, confirmed by the numerical simulations (Fig. 3e). When combined with the optical gain of InGaAsP material that is depicted by the photoluminescence (PL) spectrum (as schematically shown in Fig. 3f),  the difference in the wavelengths and $Q$s would make one of $| \sigma_\pm \rangle$ prevailing over the mode competition, to enable chiral emission of single-mode lasing under appropriate asymmetric pumping conditions. It is worth noting that, by tuning the coupling strength, the collective modes $| \sigma_\pm \rangle$ can possibly operate at the exception point (EP) region at which their frequencies are the same but $Q$s are different to support directional lasing as reported in microring lasers\cite{fengSinglemodeLaserParitytime2014,miaoOrbitalAngularMomentum2016}. However, such a specific condition is not mandatory for our design since the mode difference is already sufficient to support single-mode oscillation. More discussions are presented in Suppl. Section 3.

\section*{Sample fabrication and experimental setup}
To verify the principle and design of our vortex microlaser, we fabricate the samples from an InGaAsP MQWs wafer on the InP substrate. The PhC nanostructures are exposed by electron beam lithography (EBL) followed by inductively coupled plasma (ICP) dry etching. We subsequently remove the sacrificial layer underneath the MQWs by using hydrochloric acid to restore the mirror symmetry in the $z$ direction required by the BICs (See Methods for details). The fabricated sample is observed by using the scanning electron microscope (SEM). The top view (Fig. 4a) shows that the microlaser has a total footprint of $26.8 \times 26.8 ~\mu$m$^2$, in which the regions A and B are separated by the circular dashed line. We further cleave a part of the PhC by a focused ion beam (FIB), to observe the detailed top and side views as presented in Fig. 4b and c, respectively. The structural parameters of PhC are characterized as the periodicity of $a=537$ nm, diameter of $r_{A}=164$ nm, and height of $h=622$ nm, matching well with our design. 

The schematic of the measurement setup for vortex beam observation is shown in Fig. 4d. The sample is optically pumped at room temperature by using a pulsed laser at the wavelength of 1064 nm, with a repetition rate of 10 kHz and a pulse duration of 2 ns. Before shining upon the sample,  the pump laser beam is tailored to be transversely asymmetric by an iris diaphragm, to enable the asymmetric pumping condition required by single-mode lasing. Further, the pump beam is attenuated by an absorptive neutral density (ND) filter with an attenuation ratio of -24 dB to avoid sample damage. Accordingly, the average pump power is measured by a power meter before the ND filter, to minimize power fluctuation and noise disturbance, particularly when the pump power is low. 

The sample reaches lasing oscillation when a sufficiently strong pump beam is focused on the sample surface by an objective lens (X50), consequently, generating a vortex beam emitting in the vertical direction. The vortex and reflected pump beams are collected by the same objective lens and then pass through a lens (L1) to form a real space image at its focal point. Further, the real space image is enlarged 6 times by a 4$f$ system composed of the lenses L2 and L3, following a long-pass filter with a cut-on wavelength of 1300 nm to get rid of the reflected pump beam. Finally, the vortex beam is split and recombined to form a Mach-Zehnder interferometer for off-center self-interference observation, which is composed of two beam splitters (BSs) and two gold-coated mirrors (M1 and M2). The optical path difference is controlled by adjusting the attitude angle of M2. The real-space image of the vortex beam is captured by the complementary metal oxide semiconductor (CMOS) camera, enabling the characterizing of its real-space pattern, polarization distribution, as well as self-interference fringes to identify the expected spiral phase features. At the same time, the emission spectrum is detected by a monochromator in the range from 1540 nm to 1570 nm with a resolution of $\sim$ 0.08 nm. See Methods for more details about the measurement system.

\section*{Experimental results}
To achieve lasing oscillation, we first apply a symmetric pump beam with a circular spot (upper panel, Fig. 5a) whose diameter is about $5~\mu \mathrm{m}$. At a low pump power of 16 kW/cm$^2$, spontaneous emission is notable, along with two small peaks near the wavelength of 1555 nm, corresponding to the CW and CCW collective modes that compete in lasing. By increasing the pump power to 18 kW/cm$^2$, lasing oscillation is observed, characterized by the distinct lasing peaks amidst the strongly suppressed spontaneous emission spectrum.  It is important to highlight that the peaks of CW and CCW modes coexist in the lasing spectrum with different but comparable magnitudes, indicating a lack of dominance in mode competition. When we further elevate the pump power to a higher value of 48 kW/cm$^2$, the emission power increases significantly, while the presence of the two peaks persists, suggesting the laser is not operating under single-mode oscillation.

The above observation aligns with our theoretical prediction, showing that spontaneous chiral symmetry breaking raised by fabrication imperfection is not adequate for one of the chiral modes to dominate the mode competition. Consequently, we opt for asymmetric pumping with an elliptical spot on the sample to deliberately break the chiral symmetry, which should give rise to different $Q$ values for the collective modes in opposite chirality as we proposed in the theory. The establishment process of lasing oscillation is similar to the case of symmetric pumping during the increasing of the pump power from 16 kW/cm$^2$ to 48 kW/cm$^2$ (lower panel, Fig. 5a).  Below the lasing threshold, we still found two peaks residing in the spontaneous emission background. Differently, only one lasing peak is evident in the spectrum when the pump power exceeds the lasing threshold, proving the suppression of another chiral mode during mode competition that facilitates single-mode lasing and vortex generation. 

Further, we measure the pump-emission power curve of the sample at room temperature (Fig. 5b), in which three data points are included (labeled A to C), each corresponding to different pump power densities with their spectra illustrated in the lower panel of Fig. 5a, respectively.
Notably, a low lasing threshold of $18~\mathrm{kW/cm^2}$ is observed, attributed to the high-$Q$ characteristics of the collective modes of quasi-BICs.  The inset in Fig. 5b displays the real-space pattern of the vortex microlaser, revealing a donut shape that agrees well with our theoretical predictions. The pattern is rotationally uniform along the azimuth direction and features one peak in the radial direction, thus corresponding to the Bessel function of $(m,l)=(1,1)$. We further insert a linear polarizer before the CMOS camera to characterize the polarization distribution of the vortex beam along the orientation of $0^\circ$, $45^\circ$,  $90^\circ$, and $135^\circ$, respectively. As shown in Fig. 5c, the pattern manifests identical densities along arbitrary directions, confirming the rotational invariability of the vortex beam. For a given polarized direction, the dominant polarization aligns along the azimuth rather than the radial direction, which is quite different from the micro-ring vortex laser. This is because the collective modes primarily propagate along the axis of the circle as transverse waves but gradually rotate their directions due to boundary scatterings.  
It is also worth noting that besides the lasing mode of  $(1,1)$, other orders of collective modes have also been observed in the experiment. However, they are less favorable for lasing since their wavelengths and $Q$s values are not optimized in purpose. More data and discussion about the vortex lasing are presented in Suppl. Section 4.

Finally, we validate the spiral phase of the vortex beam by using the self-interference technique. We apply asymmetric pumping to select one of the chiral modes for lasing. Subsequently, we split the vortex beam evenly into two parts and overlapped them in real space based on a Mach-Zehnder interferometer setup.  A pair of fork patterns with opposite orientations are observed, showing the distinctive feature of phase singularity. The dislocation points of these forks align with the center of each vortex beam, at which a single fringe separates into two individual branches, confirming the vortex beam's quantum number of $|m| =  1$. Notably, the orientation of the fork pattern signifies the chirality of the lasing mode, which can be either CW or CCW, depending on how chiral symmetry is broken. As a result, we ascertain that the data shown in Fig. 5d represent the lasing of the CW mode, while we also observe the lasing of the CCW mode in another sample with the fork orientation flipped. It is worth noting that phase vortex, polarization vortex, and their combination can exhibit different features in self-interference patterns. For additional data and discussion, see Suppl. Section 5 for the details.

\section*{Discussion}
Collective modes are ubiquitous phenomena in photonic \cite{ miroshnichenkoFanoResonancesAllDielectric2012,kuznetsovOpticallyResonantDielectric2016,limonovFanoResonancesPhotonics2017}, plasmonic \cite{yanEndCentralPlasmon2007,auguieCollectiveResonancesGold2008, gianniniLightingMultipolarSurface2010}, and quantum wave systems \cite{daiTwoDimensionalDoubleQuantumSpectra2012, machaImplementationQuantumMetamaterial2014}, which refer to a cluster of individual modes collectively oscillating as a whole but showing distinctive differences in their field distribution and $Q$s than the single ones. The key to creating collective modes is to induce couplings between the 
individual modes. Although most reported collective modes are realized by coupling tight-binding modes in real space to enable collective behavior in momentum space, we propose that a similar philosophy also holds for the momentum-space localized mode (such as GMRs) to generate exotic distributions in real space, according to the duality between the real and momentum spaces.  We anticipate that, beyond boundary scatterings, other mechanisms such as random disorders, defects, and superlattices can also contribute to the couplings between single individual modes even in non-Hermitian or non-Abelian manners, thus leading to rich and unexplored phenomena of collective oscillation, showing as a direct example in photonic to reflect the famous remarks of ``more is different".

From a technical perspective, the collective modes of GMRs could be a promising way for on-chip vortex generation towards practical applications. Firstly, compared to the micro-ring vortex laser\cite{miaoOrbitalAngularMomentum2016,zhangTunableTopologicalCharge2020,zhangUltrafastControlFractional2020,chenBrightSolidstateSources2021, zhangSpinOrbitMicrolaser2022}, our approach offers a lower lasing threshold due to the high-$Q$ characteristics of quasi-BICs. Moreover, its surface-emitting nature\cite{kodigalaLasingActionPhotonic2017,gaoDiracvortexTopologicalCavities2020,yangTopologicalcavitySurfaceemittingLaser2022, contractorScalableSinglemodeSurfaceemitting2022,luanReconfigurableMoireNanolaser2023} can facilitate sufficiently high output power, which is crucial for optical manipulation, detection, and communication.  Secondly, the proposed architecture is readily compatible with electrical pumping \cite{hiroseWattclassHighpowerHighbeamquality2014, yoshidaDoublelatticePhotoniccrystalResonators2019, yoshidaHighbrightnessScalableContinuouswave2023}, which is an important stride toward practical lasers. This feasibility arises because the strict $z$-mirror symmetry required by off-$\Gamma$ BICs can be relaxed if quasi-BICs near a single symmetric-protected BIC can give sufficient high $Q$s to sustain lasing oscillation, therefore the vortex microlaser can be fabricated by using standard semiconductor laser process. 

\section*{Conclusion}
In summary, we propose and demonstrate an on-chip vortex microlaser by utilizing the collective modes of GMRs.  Leveraging the scatterings from circular boundaries, a succession of bulk GMRs that are protected by quasi-BICs collectively oscillate in real space and feature a spiral phase front in their radiation. We achieve a low lasing threshold of $18~\mathrm{kW/cm^2}$ at room temperature, and further, observe single-mode lasing behavior attributed to the spontaneous breaking of chiral symmetry under fabrication imperfections and asymmetric pumping. The chiral emission nature of the lasing beam is characterized by using polarization-resolve imaging and self-interference technique, confirming the presence of the spiral phase front. Our findings demonstrate that the coupling effect in momentum space enables the emergence of exotic collective modes in real space, thus offering a promising approach for on-chip vortex generation to fascinate many applications ranging from optical imaging, sensing, and manipulating, to large-volume optical communication.

% Your references go at the end of the main text, and before the
% figures.  For this document we've used BibTeX, the .bib file
% scibib.bib, and the .bst file Science.bst.  The package scicite.sty
% was included to format the reference numbers according to *Science*
% style.

%BibTeX users: After compilation, comment out the following two lines and paste in
% the generated .bbl file. 

%Here you should list the contents of your Supplementary Materials -- below is an example. 
%You should include a list of Supplementary figures, Tables, and any references that appear only in the SM. 
%Note that the reference numbering continues from the main text to the SM.
% In the example below, Refs. 4-10 were cited only in the SM.     
\section*{Supplementary materials}
\section*{Methods}
\section*{\underline{Numerical simulations}}
The numerical simulation is based on the finite-element method (FEM, COMSOL Multiphysics). The bandstructures in Fig.1b and the quality factors of TE-A band in Fig. 3a and b, are calculated in the 3D unit-cell structures with Floquet periodic boundaries on the sidewalls and perfectly matched layers on both top and bottom. Fig.2 and Fig.3c are performed based on 3D simulations of our microlaser design and the direct results obtained by simulations are two-fold degenerate eigenmodes independent of time. We remix the two-fold degenerate eigenmodes with a ratio of $1: e^{\pm i\frac{\pi}{2}}$ to generate CW or CCW modes. Though different remixing ratios are also permitted theoretically, these cases can be regarded as mode splitting due to asymmetric pumping. More details are presented in Suppl. Section 3.

\section*{\underline{Sample fabrication}}
The vortex microlaser with undercut PhC structure is designed on an InP-based epi-wafer. As shown in Fig.S1a, the active region consisting of six $7.5$ nm compressively strained InGaAsP quantum wells and seven $12$ nm lattice-matched InGaAsP barriers, is sandwiched by undoped $246.5$ nm InGaAsP cladding layers. An undoped $1500$ nm InP is used as wet-etching layer for the fabrication of undercut structure, leading to $z$-direction symmetry for the whole PhC. An undoped $100$ nm InGaAsP underneath the etching layer is designed as etching stop layer. The PhC structures locate at the top most of the epi-wafer. A circular central region with a radius of $10 \times a$ (periodicity $a = 537$ nm, and radius $r_{A} = 164 $ nm) is surrounded by a $50 \times 50$ array of square latticed circular holes (periodicity $a = 537$ nm, and radius $r_{B} = 154 $ nm). The PhC structures are patterned by EBL and inductively coupled plasma (ICP) in \ce{Cl2} at $240^{\circ}$C. $1:1=$ \ce{HCl} $:$ \ce{H2O} solution is used to etch InP for the undercut structure under a wet-etching rate of $700$ nm$/$min at room temperature.

\section*{\underline{Measurement and data processing}}
We use a 1064-nm pulsed laser with a repetition frequency of 10 kHz and a pulse duration of 2 ns, to pump our vortex microlaser at room temperature. As shown in Fig.4d, a $\times50$ objective lens (IOPAMI137150X-NIR, Mitutoyo) is used to converge the pump light to a spot with the diameter of $\sim 5$ um and also collect the vortex beam generated from the sample. The following lens L1 ($f = 200~\mathrm{mm}$) confocal with the objective lens, as well as L2 ($f = 50~\mathrm{mm}$) and L3 ($f = 300~\mathrm{mm}$) forming a 4$f$ system, enlarges the real space image of samples by 300 times in total to generate more in-plane optical path difference in the self-interference experiment. The real space images are captured by an InGaAs infrared CMOS camera (IMX990, Sony). Meanwhile, they are also focused into a monochromator with a 600g/mm optical grating and an infrared array detector (IsoPlane SCT320 and NIRvana 640, Princeton Instruments). The images, after deducting CMOS intrinsic noise and ambient background noise, are applied data-smoothing with a window size of 30 pixels.

\section*{References}
\bibliography{mainbib}

\begin{thebibliography}{10}
\expandafter\ifx\csname url\endcsname\relax
  \def\url#1{\texttt{#1}}\fi
\expandafter\ifx\csname urlprefix\endcsname\relax\def\urlprefix{URL }\fi
\providecommand{\bibinfo}[2]{#2}
\providecommand{\eprint}[2][]{\url{#2}}

\bibitem{allenOrbitalAngularMomentum1992}
\bibinfo{author}{Allen, L.}, \bibinfo{author}{Beijersbergen, M.~W.}, \bibinfo{author}{Spreeuw, R. J.~C.} \& \bibinfo{author}{Woerdman, J.~P.}
\newblock \bibinfo{title}{Orbital angular momentum of light and the transformation of {{Laguerre-Gaussian}} laser modes}.
\newblock \emph{\bibinfo{journal}{Phys. Rev. A}} \textbf{\bibinfo{volume}{45}}, \bibinfo{pages}{8185--8189}
\newblock  (\bibinfo{year}{1992}).

\bibitem{laveryDetectionSpinningObject2013}
\bibinfo{author}{Lavery, M. P.~J.}, \bibinfo{author}{Speirits, F.~C.}, \bibinfo{author}{Barnett, S.~M.} \& \bibinfo{author}{Padgett, M.~J.}
\newblock \bibinfo{title}{Detection of a {{Spinning Object Using Light}}'s {{Orbital Angular Momentum}}}.
\newblock \emph{\bibinfo{journal}{Science}} \textbf{\bibinfo{volume}{341}}, \bibinfo{pages}{537--540}
\newblock  (\bibinfo{year}{2013}).

\bibitem{patersonControlledRotationOptically2001}
\bibinfo{author}{Paterson, L.} \emph{et~al.}
\newblock \bibinfo{title}{Controlled {{Rotation}} of {{Optically Trapped Microscopic Particles}}}.
\newblock \emph{\bibinfo{journal}{Science}} \textbf{\bibinfo{volume}{292}}, \bibinfo{pages}{912--914}
\newblock  (\bibinfo{year}{2001}).

\bibitem{macdonaldCreationManipulationThreeDimensional2002}
\bibinfo{author}{MacDonald, M.~P.} \emph{et~al.}
\newblock \bibinfo{title}{Creation and {{Manipulation}} of {{Three-Dimensional Optically Trapped Structures}}}.
\newblock \emph{\bibinfo{journal}{Science}} \textbf{\bibinfo{volume}{296}}, \bibinfo{pages}{1101--1103}
\newblock  (\bibinfo{year}{2002}).

\bibitem{wangTerabitFreespaceData2012}
\bibinfo{author}{Wang, J.} \emph{et~al.}
\newblock \bibinfo{title}{Terabit free-space data transmission employing orbital angular momentum multiplexing}.
\newblock \emph{\bibinfo{journal}{Nat. Photon.}} \textbf{\bibinfo{volume}{6}}, \bibinfo{pages}{488--496}
\newblock  (\bibinfo{year}{2012}).

\bibitem{leiMassiveIndividualOrbital2015}
\bibinfo{author}{Lei, T.} \emph{et~al.}
\newblock \bibinfo{title}{Massive individual orbital angular momentum channels for multiplexing enabled by {{Dammann}} gratings}.
\newblock \emph{\bibinfo{journal}{Light Sci. Appl.}} \textbf{\bibinfo{volume}{4}}, \bibinfo{pages}{e257--e257}
\newblock  (\bibinfo{year}{2015}).

\bibitem{xieUltrabroadbandOnchipTwisted2018}
\bibinfo{author}{Xie, Z.} \emph{et~al.}
\newblock \bibinfo{title}{Ultra-broadband on-chip twisted light emitter for optical communications}.
\newblock \emph{\bibinfo{journal}{Light Sci. Appl.}} \textbf{\bibinfo{volume}{7}}, \bibinfo{pages}{18001--18001}
\newblock  (\bibinfo{year}{2018}).

\bibitem{chenIntegratedOpticalVortex2024}
\bibinfo{author}{Chen, B.} \emph{et~al.}
\newblock \bibinfo{title}{Integrated optical vortex microcomb}.
\newblock \emph{\bibinfo{journal}{Nat. Photon.}}
\newblock  (\bibinfo{year}{2024}).

\bibitem{caiIntegratedCompactOptical2012}
\bibinfo{author}{Cai, X.} \emph{et~al.}
\newblock \bibinfo{title}{Integrated {{Compact Optical Vortex Beam Emitters}}}.
\newblock \emph{\bibinfo{journal}{Science}} \textbf{\bibinfo{volume}{338}}, \bibinfo{pages}{363--366}
\newblock  (\bibinfo{year}{2012}).

\bibitem{liuDirectFiberVector2017}
\bibinfo{author}{Liu, J.} \emph{et~al.}
\newblock \bibinfo{title}{Direct fiber vector eigenmode multiplexing transmission seeded by integrated optical vortex emitters}.
\newblock \emph{\bibinfo{journal}{Light Sci. Appl.}} \textbf{\bibinfo{volume}{7}}, \bibinfo{pages}{17148--17148}
\newblock  (\bibinfo{year}{2017}).

\bibitem{shaoSpinorbitInteractionLight2018}
\bibinfo{author}{Shao, Z.}, \bibinfo{author}{Zhu, J.}, \bibinfo{author}{Chen, Y.}, \bibinfo{author}{Zhang, Y.} \& \bibinfo{author}{Yu, S.}
\newblock \bibinfo{title}{Spin-orbit interaction of light induced by transverse spin angular momentum engineering}.
\newblock \emph{\bibinfo{journal}{Nat. Commun.}} \textbf{\bibinfo{volume}{9}}, \bibinfo{pages}{926}
\newblock  (\bibinfo{year}{2018}).

\bibitem{zhouUltracompactBroadbandPolarization2019}
\bibinfo{author}{Zhou, N.} \emph{et~al.}
\newblock \bibinfo{title}{Ultra-compact broadband polarization diversity orbital angular momentum generator with 3.6 {\texttimes} 3.6 {$M$}m2 footprint}.
\newblock \emph{\bibinfo{journal}{Sci. Adv.}} \textbf{\bibinfo{volume}{5}}, \bibinfo{pages}{eaau9593}
\newblock  (\bibinfo{year}{2019}).

\bibitem{sroorHighpurityOrbitalAngular2020}
\bibinfo{author}{Sroor, H.} \emph{et~al.}
\newblock \bibinfo{title}{High-purity orbital angular momentum states from a visible metasurface laser}.
\newblock \emph{\bibinfo{journal}{Nat. Photon.}} \textbf{\bibinfo{volume}{14}}, \bibinfo{pages}{498--503}
\newblock  (\bibinfo{year}{2020}).

\bibitem{wangVortexRadiationSingle2022}
\bibinfo{author}{Wang, X.-Y.} \emph{et~al.}
\newblock \bibinfo{title}{Vortex radiation from a single emitter in a chiral plasmonic nanocavity}.
\newblock \emph{\bibinfo{journal}{Nanophotonics}} \textbf{\bibinfo{volume}{11}}, \bibinfo{pages}{1905--1911}
\newblock  (\bibinfo{year}{2022}).

\bibitem{meiCascadedMetasurfacesHighpurity2023}
\bibinfo{author}{Mei, F.} \emph{et~al.}
\newblock \bibinfo{title}{Cascaded metasurfaces for high-purity vortex generation}.
\newblock \emph{\bibinfo{journal}{Nat. Commun.}} \textbf{\bibinfo{volume}{14}}, \bibinfo{pages}{6410}
\newblock  (\bibinfo{year}{2023}).

\bibitem{liuIntegratedVortexSoliton2024}
\bibinfo{author}{Liu, Y.} \emph{et~al.}
\newblock \bibinfo{title}{Integrated vortex soliton microcombs}.
\newblock \emph{\bibinfo{journal}{Nat. Photon.}} \textbf{\bibinfo{volume}{18}}, \bibinfo{pages}{632--637}
\newblock  (\bibinfo{year}{2024}).

\bibitem{liOrbitalAngularMomentum2015}
\bibinfo{author}{Li, H.} \emph{et~al.}
\newblock \bibinfo{title}{Orbital angular momentum vertical-cavity surface-emitting lasers}.
\newblock \emph{\bibinfo{journal}{Optica}} \textbf{\bibinfo{volume}{2}}, \bibinfo{pages}{547--552}
\newblock  (\bibinfo{year}{2015}).

\bibitem{miaoOrbitalAngularMomentum2016}
\bibinfo{author}{Miao, P.} \emph{et~al.}
\newblock \bibinfo{title}{Orbital angular momentum microlaser}.
\newblock \emph{\bibinfo{journal}{Science}} \textbf{\bibinfo{volume}{353}}, \bibinfo{pages}{464--467}
\newblock  (\bibinfo{year}{2016}).

\bibitem{carlonzambonOpticallyControllingEmission2019}
\bibinfo{author}{Carlon~Zambon, N.} \emph{et~al.}
\newblock \bibinfo{title}{Optically controlling the emission chirality of microlasers}.
\newblock \emph{\bibinfo{journal}{Nat. Photon.}} \textbf{\bibinfo{volume}{13}}, \bibinfo{pages}{283--288}
\newblock  (\bibinfo{year}{2019}).

\bibitem{sunLeadHalidePerovskite2020}
\bibinfo{author}{Sun, W.} \emph{et~al.}
\newblock \bibinfo{title}{Lead halide perovskite vortex microlasers}.
\newblock \emph{\bibinfo{journal}{Nat. Commun.}} \textbf{\bibinfo{volume}{11}}, \bibinfo{pages}{4862}
\newblock  (\bibinfo{year}{2020}).

\bibitem{yangSpinMomentumLockedEdgeMode2020}
\bibinfo{author}{Yang, Z.-Q.}, \bibinfo{author}{Shao, Z.-K.}, \bibinfo{author}{Chen, H.-Z.}, \bibinfo{author}{Mao, X.-R.} \& \bibinfo{author}{Ma, R.-M.}
\newblock \bibinfo{title}{Spin-{{Momentum-Locked Edge Mode}} for {{Topological Vortex Lasing}}}.
\newblock \emph{\bibinfo{journal}{Phys. Rev. Lett.}} \textbf{\bibinfo{volume}{125}}, \bibinfo{pages}{013903}
\newblock  (\bibinfo{year}{2020}).

\bibitem{zhangTunableTopologicalCharge2020}
\bibinfo{author}{Zhang, Z.} \emph{et~al.}
\newblock \bibinfo{title}{Tunable topological charge vortex microlaser}.
\newblock \emph{\bibinfo{journal}{Science}} \textbf{\bibinfo{volume}{368}}, \bibinfo{pages}{760--763}
\newblock  (\bibinfo{year}{2020}).

\bibitem{zhangUltrafastControlFractional2020}
\bibinfo{author}{Zhang, Z.} \emph{et~al.}
\newblock \bibinfo{title}{Ultrafast control of fractional orbital angular momentum of microlaser emissions}.
\newblock \emph{\bibinfo{journal}{Light Sci. Appl.}} \textbf{\bibinfo{volume}{9}}, \bibinfo{pages}{179}
\newblock  (\bibinfo{year}{2020}).

\bibitem{chenBrightSolidstateSources2021}
\bibinfo{author}{Chen, B.} \emph{et~al.}
\newblock \bibinfo{title}{Bright solid-state sources for single photons with orbital angular momentum}.
\newblock \emph{\bibinfo{journal}{Nat. Nanotechnol.}} \textbf{\bibinfo{volume}{16}}, \bibinfo{pages}{302--307}
\newblock  (\bibinfo{year}{2021}).

\bibitem{piccardoVortexLaserArrays2022}
\bibinfo{author}{Piccardo, M.} \emph{et~al.}
\newblock \bibinfo{title}{Vortex laser arrays with topological charge control and self-healing of defects}.
\newblock \emph{\bibinfo{journal}{Nat. Photon.}} \textbf{\bibinfo{volume}{16}}, \bibinfo{pages}{359--365}
\newblock  (\bibinfo{year}{2022}).

\bibitem{zhangSpinOrbitMicrolaser2022}
\bibinfo{author}{Zhang, Z.} \emph{et~al.}
\newblock \bibinfo{title}{Spin--orbit microlaser emitting in a four-dimensional {{Hilbert}} space}.
\newblock \emph{\bibinfo{journal}{Nature}} \textbf{\bibinfo{volume}{612}}, \bibinfo{pages}{246--251}
\newblock  (\bibinfo{year}{2022}).

\bibitem{huangUltrafastControlVortex2020}
\bibinfo{author}{Huang, C.} \emph{et~al.}
\newblock \bibinfo{title}{Ultrafast control of vortex microlasers}.
\newblock \emph{\bibinfo{journal}{Science}} \textbf{\bibinfo{volume}{367}}, \bibinfo{pages}{1018--1021}
\newblock  (\bibinfo{year}{2020}).

\bibitem{mohamedControllingTopologyPolarization2022}
\bibinfo{author}{Mohamed, S.} \emph{et~al.}
\newblock \bibinfo{title}{Controlling {{Topology}} and {{Polarization State}} of {{Lasing Photonic Bound States}} in {{Continuum}}}.
\newblock \emph{\bibinfo{journal}{Laser \& Photon. Rev.}} \textbf{\bibinfo{volume}{16}}, \bibinfo{pages}{2100574}
\newblock  (\bibinfo{year}{2022}).

\bibitem{hwangVortexNanolaserBased2023}
\bibinfo{author}{Hwang, M.-S.} \emph{et~al.}
\newblock \bibinfo{title}{Vortex nanolaser based on a photonic disclination cavity}.
\newblock \emph{\bibinfo{journal}{Nat. Photon.}} \textbf{\bibinfo{volume}{18}}, \bibinfo{pages}{1--8}
\newblock  (\bibinfo{year}{2023}).

\bibitem{hsuObservationTrappedLight2013}
\bibinfo{author}{Hsu, C.~W.} \emph{et~al.}
\newblock \bibinfo{title}{Observation of trapped light within the radiation continuum}.
\newblock \emph{\bibinfo{journal}{Nature}} \textbf{\bibinfo{volume}{499}}, \bibinfo{pages}{188--191}
\newblock  (\bibinfo{year}{2013}).

\bibitem{zhenTopologicalNatureOptical2014}
\bibinfo{author}{Zhen, B.}, \bibinfo{author}{Hsu, C.~W.}, \bibinfo{author}{Lu, L.}, \bibinfo{author}{Stone, A.~D.} \& \bibinfo{author}{Solja{\v c}i{\'c}, M.}
\newblock \bibinfo{title}{Topological {{Nature}} of {{Optical Bound States}} in the {{Continuum}}}.
\newblock \emph{\bibinfo{journal}{Phys. Rev. Lett.}} \textbf{\bibinfo{volume}{113}}, \bibinfo{pages}{257401}
\newblock  (\bibinfo{year}{2014}).

\bibitem{jinTopologicallyEnabledUltrahighQ2019}
\bibinfo{author}{Jin, J.} \emph{et~al.}
\newblock \bibinfo{title}{Topologically enabled ultrahigh-{{Q}} guided resonances robust to out-of-plane scattering}.
\newblock \emph{\bibinfo{journal}{Nature}} \textbf{\bibinfo{volume}{574}}, \bibinfo{pages}{501--504}
\newblock  (\bibinfo{year}{2019}).

\bibitem{auguieCollectiveResonancesGold2008}
\bibinfo{author}{Augui{\'e}, B.} \& \bibinfo{author}{Barnes, W.~L.}
\newblock \bibinfo{title}{Collective {{Resonances}} in {{Gold Nanoparticle Arrays}}}.
\newblock \emph{\bibinfo{journal}{Phys. Rev. Lett.}} \textbf{\bibinfo{volume}{101}}, \bibinfo{pages}{143902}
\newblock  (\bibinfo{year}{2008}).

\bibitem{kuznetsovOpticallyResonantDielectric2016}
\bibinfo{author}{Kuznetsov, A.~I.}, \bibinfo{author}{Miroshnichenko, A.~E.}, \bibinfo{author}{Brongersma, M.~L.}, \bibinfo{author}{Kivshar, Y.~S.} \& \bibinfo{author}{Luk'yanchuk, B.}
\newblock \bibinfo{title}{Optically resonant dielectric nanostructures}.
\newblock \emph{\bibinfo{journal}{Science}} \textbf{\bibinfo{volume}{354}}, \bibinfo{pages}{aag2472}
\newblock  (\bibinfo{year}{2016}).

\bibitem{limonovFanoResonancesPhotonics2017}
\bibinfo{author}{Limonov, M.~F.}, \bibinfo{author}{Rybin, M.~V.}, \bibinfo{author}{Poddubny, A.~N.} \& \bibinfo{author}{Kivshar, Y.~S.}
\newblock \bibinfo{title}{Fano resonances in photonics}.
\newblock \emph{\bibinfo{journal}{Nat. Photon.}} \textbf{\bibinfo{volume}{11}}, \bibinfo{pages}{543--554}
\newblock  (\bibinfo{year}{2017}).

\bibitem{yesilkoyUltrasensitiveHyperspectralImaging2019}
\bibinfo{author}{Yesilkoy, F.} \emph{et~al.}
\newblock \bibinfo{title}{Ultrasensitive hyperspectral imaging and biodetection enabled by dielectric metasurfaces}.
\newblock \emph{\bibinfo{journal}{Nat. Photon.}} \textbf{\bibinfo{volume}{13}}, \bibinfo{pages}{390--396}
\newblock  (\bibinfo{year}{2019}).

\bibitem{hoangCollectiveNatureHighQ2024}
\bibinfo{author}{Hoang, T.~X.} \emph{et~al.}
\newblock \bibinfo{title}{Collective nature of high-{{Q}} resonances in finite-size photonic metastructures}.
\newblock \emph{\bibinfo{journal}{arXiv}} \bibinfo{pages}{2405.01034}
\newblock  (\bibinfo{year}{2024}).

\bibitem{chenObservationMiniaturizedBound2022}
\bibinfo{author}{Chen, Z.} \emph{et~al.}
\newblock \bibinfo{title}{Observation of miniaturized bound states in the continuum with ultra-high quality factors}.
\newblock \emph{\bibinfo{journal}{Sci. Bull.}} \textbf{\bibinfo{volume}{67}}, \bibinfo{pages}{359--366}
\newblock  (\bibinfo{year}{2022}).

\bibitem{notomiTheoryLightPropagation2000}
\bibinfo{author}{Notomi, M.}
\newblock \bibinfo{title}{Theory of light propagation in strongly modulated photonic crystals: {{Refractionlike}} behavior in the vicinity of the photonic band gap}.
\newblock \emph{\bibinfo{journal}{Phys. Rev. B}} \textbf{\bibinfo{volume}{62}}, \bibinfo{pages}{10696--10705}
\newblock  (\bibinfo{year}{2000}).

\bibitem{notomiNegativeRefractionPhotonic2002}
\bibinfo{author}{Notomi, M.}
\newblock \bibinfo{title}{Negative refraction in photonic crystals}.
\newblock \emph{\bibinfo{journal}{Opt. Quantum Electron.}} \textbf{\bibinfo{volume}{34}}, \bibinfo{pages}{133--143}
\newblock  (\bibinfo{year}{2002}).

\bibitem{liangThreedimensionalCoupledwaveModel2011}
\bibinfo{author}{Liang, Y.}, \bibinfo{author}{Peng, C.}, \bibinfo{author}{Sakai, K.}, \bibinfo{author}{Iwahashi, S.} \& \bibinfo{author}{Noda, S.}
\newblock \bibinfo{title}{Three-dimensional coupled-wave model for square-lattice photonic crystal lasers with transverse electric polarization: {{A}} general approach}.
\newblock \emph{\bibinfo{journal}{Phys. Rev. B}} \textbf{\bibinfo{volume}{84}}, \bibinfo{pages}{195119}
\newblock  (\bibinfo{year}{2011}).

\bibitem{chenAnalyticalTheoryFinitesize2022}
\bibinfo{author}{Chen, Z.} \emph{et~al.}
\newblock \bibinfo{title}{Analytical theory of finite-size photonic crystal slabs near the band edge}.
\newblock \emph{\bibinfo{journal}{Opt. Express}} \textbf{\bibinfo{volume}{30}}, \bibinfo{pages}{14033--14047}
\newblock  (\bibinfo{year}{2022}).

\bibitem{renLowthresholdNanolasersBased2022}
\bibinfo{author}{Ren, Y.} \emph{et~al.}
\newblock \bibinfo{title}{Low-threshold nanolasers based on miniaturized bound states in the continuum}.
\newblock \emph{\bibinfo{journal}{Sci. Adv.}} \textbf{\bibinfo{volume}{8}}, \bibinfo{pages}{eade8817}
\newblock  (\bibinfo{year}{2022}).

\bibitem{wiersigAsymmetricScatteringNonorthogonal2008}
\bibinfo{author}{Wiersig, J.}, \bibinfo{author}{Kim, S.~W.} \& \bibinfo{author}{Hentschel, M.}
\newblock \bibinfo{title}{Asymmetric scattering and nonorthogonal mode patterns in optical microspirals}.
\newblock \emph{\bibinfo{journal}{Phys. Rev. A}} \textbf{\bibinfo{volume}{78}}, \bibinfo{pages}{053809}
\newblock  (\bibinfo{year}{2008}).

\bibitem{wiersigNonorthogonalPairsCopropagating2011}
\bibinfo{author}{Wiersig, J.} \emph{et~al.}
\newblock \bibinfo{title}{Nonorthogonal pairs of copropagating optical modes in deformed microdisk cavities}.
\newblock \emph{\bibinfo{journal}{Phys. Rev. A}} \textbf{\bibinfo{volume}{84}}, \bibinfo{pages}{023845}
\newblock  (\bibinfo{year}{2011}).

\bibitem{wiersigStructureWhisperinggalleryModes2011}
\bibinfo{author}{Wiersig, J.}
\newblock \bibinfo{title}{Structure of whispering-gallery modes in optical microdisks perturbed by nanoparticles}.
\newblock \emph{\bibinfo{journal}{Phys. Rev. A}} \textbf{\bibinfo{volume}{84}}, \bibinfo{pages}{063828}
\newblock  (\bibinfo{year}{2011}).

\bibitem{pengChiralModesDirectional2016}
\bibinfo{author}{Peng, B.} \emph{et~al.}
\newblock \bibinfo{title}{Chiral modes and directional lasing at exceptional points}.
\newblock \emph{\bibinfo{journal}{Proc. Natl. Acad. Sci.}} \textbf{\bibinfo{volume}{113}}, \bibinfo{pages}{6845--6850}
\newblock  (\bibinfo{year}{2016}).

\bibitem{chenExceptionalPointsEnhance2017}
\bibinfo{author}{Chen, W.}, \bibinfo{author}{Kaya~{\"O}zdemir, {\c S}.}, \bibinfo{author}{Zhao, G.}, \bibinfo{author}{Wiersig, J.} \& \bibinfo{author}{Yang, L.}
\newblock \bibinfo{title}{Exceptional points enhance sensing in an optical microcavity}.
\newblock \emph{\bibinfo{journal}{Nature}} \textbf{\bibinfo{volume}{548}}, \bibinfo{pages}{192--196}
\newblock  (\bibinfo{year}{2017}).

\bibitem{fengSinglemodeLaserParitytime2014}
\bibinfo{author}{Feng, L.}, \bibinfo{author}{Wong, Z.~J.}, \bibinfo{author}{Ma, R.-M.}, \bibinfo{author}{Wang, Y.} \& \bibinfo{author}{Zhang, X.}
\newblock \bibinfo{title}{Single-mode laser by parity-time symmetry breaking}.
\newblock \emph{\bibinfo{journal}{Science}} \textbf{\bibinfo{volume}{346}}, \bibinfo{pages}{972--975}
\newblock  (\bibinfo{year}{2014}).

\bibitem{miroshnichenkoFanoResonancesAllDielectric2012}
\bibinfo{author}{Miroshnichenko, A.~E.} \& \bibinfo{author}{Kivshar, Y.~S.}
\newblock \bibinfo{title}{Fano {{Resonances}} in {{All-Dielectric Oligomers}}}.
\newblock \emph{\bibinfo{journal}{Nano Lett.}} \textbf{\bibinfo{volume}{12}}, \bibinfo{pages}{6459--6463}
\newblock  (\bibinfo{year}{2012}).

\bibitem{yanEndCentralPlasmon2007}
\bibinfo{author}{Yan, J.}, \bibinfo{author}{Yuan, Z.} \& \bibinfo{author}{Gao, S.}
\newblock \bibinfo{title}{End and {{Central Plasmon Resonances}} in {{Linear Atomic Chains}}}.
\newblock \emph{\bibinfo{journal}{Phys. Rev. Lett.}} \textbf{\bibinfo{volume}{98}}, \bibinfo{pages}{216602}
\newblock  (\bibinfo{year}{2007}).

\bibitem{gianniniLightingMultipolarSurface2010}
\bibinfo{author}{Giannini, V.}, \bibinfo{author}{Vecchi, G.} \& \bibinfo{author}{G{\'o}mez~Rivas, J.}
\newblock \bibinfo{title}{Lighting {{Up Multipolar Surface Plasmon Polaritons}} by {{Collective Resonances}} in {{Arrays}} of {{Nanoantennas}}}.
\newblock \emph{\bibinfo{journal}{Phys. Rev. Lett.}} \textbf{\bibinfo{volume}{105}}, \bibinfo{pages}{266801}
\newblock  (\bibinfo{year}{2010}).

\bibitem{daiTwoDimensionalDoubleQuantumSpectra2012}
\bibinfo{author}{Dai, X.} \emph{et~al.}
\newblock \bibinfo{title}{Two-{{Dimensional Double-Quantum Spectra Reveal Collective Resonances}} in an {{Atomic Vapor}}}.
\newblock \emph{\bibinfo{journal}{Phys. Rev. Lett.}} \textbf{\bibinfo{volume}{108}}, \bibinfo{pages}{193201}
\newblock  (\bibinfo{year}{2012}).

\bibitem{machaImplementationQuantumMetamaterial2014}
\bibinfo{author}{Macha, P.} \emph{et~al.}
\newblock \bibinfo{title}{Implementation of a quantum metamaterial using superconducting qubits}.
\newblock \emph{\bibinfo{journal}{Nat. Commun.}} \textbf{\bibinfo{volume}{5}}, \bibinfo{pages}{5146}
\newblock  (\bibinfo{year}{2014}).

\bibitem{kodigalaLasingActionPhotonic2017}
\bibinfo{author}{Kodigala, A.} \emph{et~al.}
\newblock \bibinfo{title}{Lasing action from photonic bound states in continuum}.
\newblock \emph{\bibinfo{journal}{Nature}} \textbf{\bibinfo{volume}{541}}, \bibinfo{pages}{196--199}
\newblock  (\bibinfo{year}{2017}).

\bibitem{gaoDiracvortexTopologicalCavities2020}
\bibinfo{author}{Gao, X.} \emph{et~al.}
\newblock \bibinfo{title}{Dirac-vortex topological cavities}.
\newblock \emph{\bibinfo{journal}{Nat. Nanotechnol.}} \textbf{\bibinfo{volume}{15}}, \bibinfo{pages}{1012--1018}
\newblock  (\bibinfo{year}{2020}).

\bibitem{yangTopologicalcavitySurfaceemittingLaser2022}
\bibinfo{author}{Yang, L.}, \bibinfo{author}{Li, G.}, \bibinfo{author}{Gao, X.} \& \bibinfo{author}{Lu, L.}
\newblock \bibinfo{title}{Topological-cavity surface-emitting laser}.
\newblock \emph{\bibinfo{journal}{Nat. Photon.}} \textbf{\bibinfo{volume}{16}}, \bibinfo{pages}{279--283}
\newblock  (\bibinfo{year}{2022}).

\bibitem{contractorScalableSinglemodeSurfaceemitting2022}
\bibinfo{author}{Contractor, R.} \emph{et~al.}
\newblock \bibinfo{title}{Scalable single-mode surface-emitting laser via open-{{Dirac}} singularities}.
\newblock \emph{\bibinfo{journal}{Nature}} \textbf{\bibinfo{volume}{608}}, \bibinfo{pages}{692--698}
\newblock  (\bibinfo{year}{2022}).

\bibitem{luanReconfigurableMoireNanolaser2023}
\bibinfo{author}{Luan, H.-Y.}, \bibinfo{author}{Ouyang, Y.-H.}, \bibinfo{author}{Zhao, Z.-W.}, \bibinfo{author}{Mao, W.-Z.} \& \bibinfo{author}{Ma, R.-M.}
\newblock \bibinfo{title}{Reconfigurable moir{\'e} nanolaser arrays with phase synchronization}.
\newblock \emph{\bibinfo{journal}{Nature}} \textbf{\bibinfo{volume}{624}}, \bibinfo{pages}{282--288}
\newblock  (\bibinfo{year}{2023}).

\bibitem{hiroseWattclassHighpowerHighbeamquality2014}
\bibinfo{author}{Hirose, K.} \emph{et~al.}
\newblock \bibinfo{title}{Watt-class high-power, high-beam-quality photonic-crystal lasers}.
\newblock \emph{\bibinfo{journal}{Nat. Photon.}} \textbf{\bibinfo{volume}{8}}, \bibinfo{pages}{406--411}
\newblock  (\bibinfo{year}{2014}).

\bibitem{yoshidaDoublelatticePhotoniccrystalResonators2019}
\bibinfo{author}{Yoshida, M.} \emph{et~al.}
\newblock \bibinfo{title}{Double-lattice photonic-crystal resonators enabling high-brightness semiconductor lasers with symmetric narrow-divergence beams}.
\newblock \emph{\bibinfo{journal}{Nat. Mater.}} \textbf{\bibinfo{volume}{18}}, \bibinfo{pages}{121--128}
\newblock  (\bibinfo{year}{2019}).

\bibitem{yoshidaHighbrightnessScalableContinuouswave2023}
\bibinfo{author}{Yoshida, M.} \emph{et~al.}
\newblock \bibinfo{title}{High-brightness scalable continuous-wave single-mode photonic-crystal laser}.
\newblock \emph{\bibinfo{journal}{Nature}} \textbf{\bibinfo{volume}{618}}, \bibinfo{pages}{727--732}
\newblock  (\bibinfo{year}{2023}).

\end{thebibliography}
\bibliographystyle{naturemag}

% For your review copy (i.e., the file you initially send in for
% evaluation), you can use the {figure} environment and the
% \includegraphics command to stream your figures into the text, placing
% all figures at the end.  For the final, revised manuscript for
% acceptance and production, however, PostScript or other graphics
% should not be streamed into your compliled file.  Instead, set
% captions as simple paragraphs (with a \noindent tag), setting them
% off from the rest of the text with a \clearpage as shown  below, and
% submit figures as separate files according to the Art Department's
% instructions.

\clearpage

\begin{figure}[htbp] 
 \centering 
 \includegraphics[width=16cm]{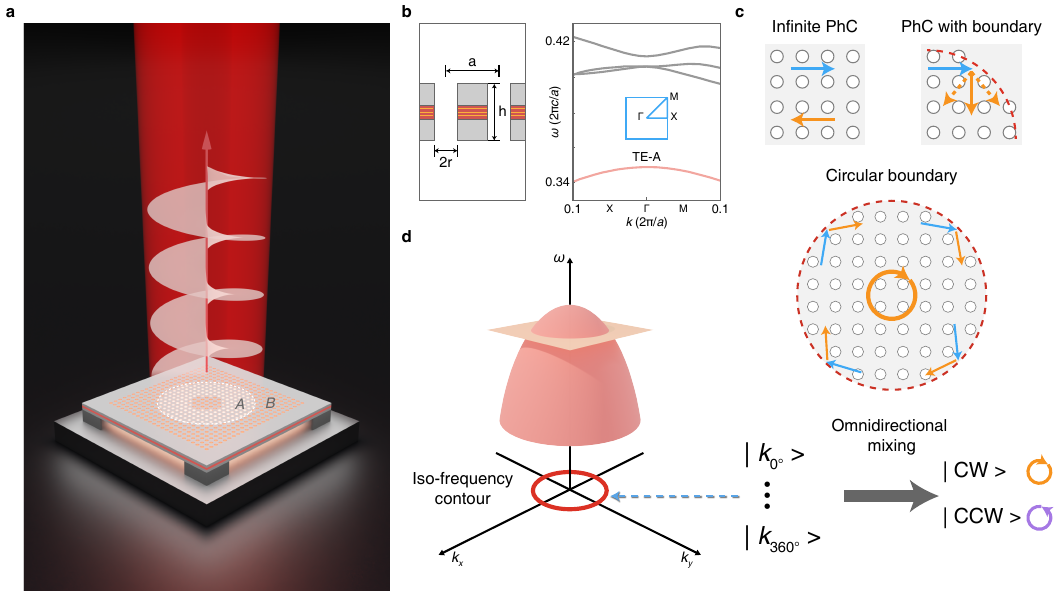}
\caption{\textbf{Principle of collective modes}
(a)	
Schematic of the microlaser showing vortex generation under optical pumping. The central PhC region A is surrounded by a heterogeneous PhC region B with a circular boundary.
(b)	
The PhC geometry is defined by lattice constant $a$, air-hole radius $r$, and slab thickness $h$  (left panel), which gives rise to a band structure accordingly (right panel), where the TE-A band of region A near the 2nd-$\Gamma$ point is embedded in the bandgap of region B. 
(c)
The principle of boundary scattering. The Bloch waves propagate along straight directions in an infinite periodic PhC (left upper panel), while the boundary scatterings give additional momenta $\Delta k$ to alter their directions (right upper panel). As a result, an omnidirectional mixing of Bloch waves happens at the ideal circular boundary (lower panel). 
(d)
A 3D visualization of the TE-A band's dispersion, showing an isotropic iso-frequency contour at the Brillouin zone center to support the collective modes.}

\label{Fig1}
 
\end{figure}

\clearpage
\begin{figure}[htbp] 
 \centering 
 \includegraphics[width=16cm]{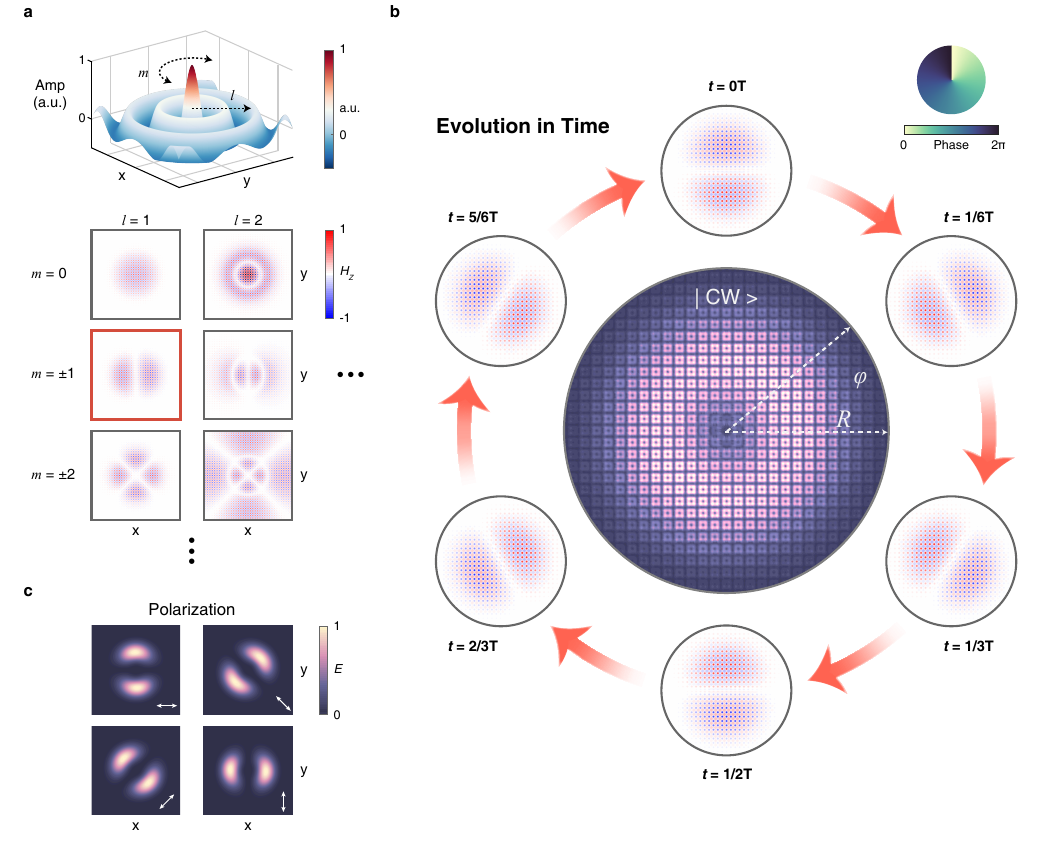}
\caption{\textbf{Spiral phase in the real space }
(a)
The amplitudes of collective modes in our circular PhC cavity are Bessel functions $J_m(r)$ with quantum numbers $m$ and $l$ in the azimuth and radial directions, respectively (upper panel), and their $H_z$ field distributions are plotted (lower panel). 
(b)
As the candidate for lasing oscillation, the field strength ($|E|$) of collective mode $(1,1)$ is illustrated as a donut pattern. Moreover, the snapshots of magnetic fields $H_z$ at a sequence of time intervals in one oscillation cycle $[0, T]$ show the temporal rotating of the mode that generates a spiral phase from $0$ to $2\pi$. 
(c)
The polarization-solved distribution of collective mode $(1,1)$ exhibits two lobes with equal density intensities as a distinct feature of the vortex beam. }
\label{Fig2}
\end{figure}

\clearpage
\begin{figure}[htbp] 
 \centering 
 \includegraphics[width=12cm]{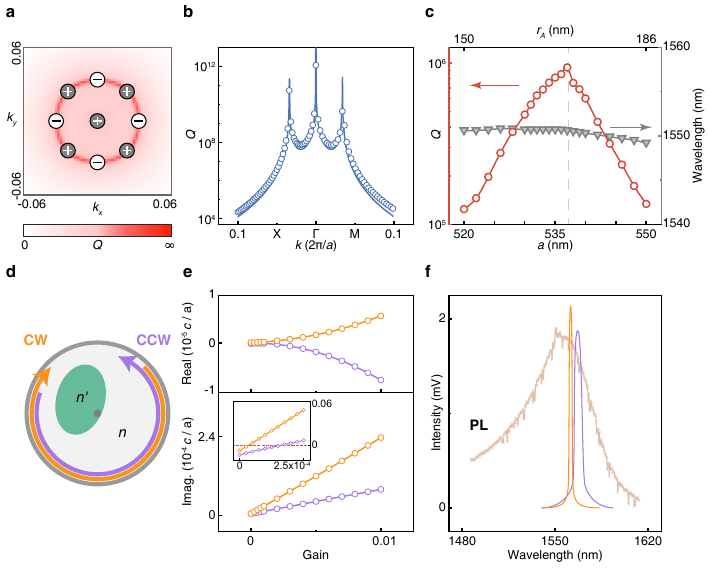}
\caption{\textbf{High-$Q$ and chiral nature of  collective modes}
(a) The $Q$ distribution in momentum space for bulk GMRs in the PhC slab, showing eight off-$\Gamma$ BICs and one symmetry-protected BIC appear near the Brillouin zone center, represented by integer topological charges where the $Q$s diverge to infinite. 
(b)
The detailed $Q$s of the TE-A band, the off-$\Gamma$ BICs are found at $|k_{\text{BIC}}|=0.034$.
(c)
The $Q$s of collective mode $(1,1)$ in the PhC with circular boundary, the $Q$s are optimized by tuning structural parameters while keeping the wavelength almost fixed at  $\sim1550$, the maximum $Q$$\sim 1\times 10^6$ appears at $a=537$ nm. 
(d)
Under asymmetric pumping, the CW and CCW collective modes behave differently under an elliptical pump spot (blue).
(e) The real and imaginary eigenfrequencies under asymmetric pumping, showing differences in the wavelengths and $Q$s for mode competition. 
(f)
The principle of single-mode oscillation. The differences of CW and CCW collective modes upon the PL gain specturm enable one chiral mode to prevail in lasing.}
\label{Fig3}
\end{figure}

\clearpage
\begin{figure}[htbp] 
 \centering 
 \includegraphics[width=16cm]{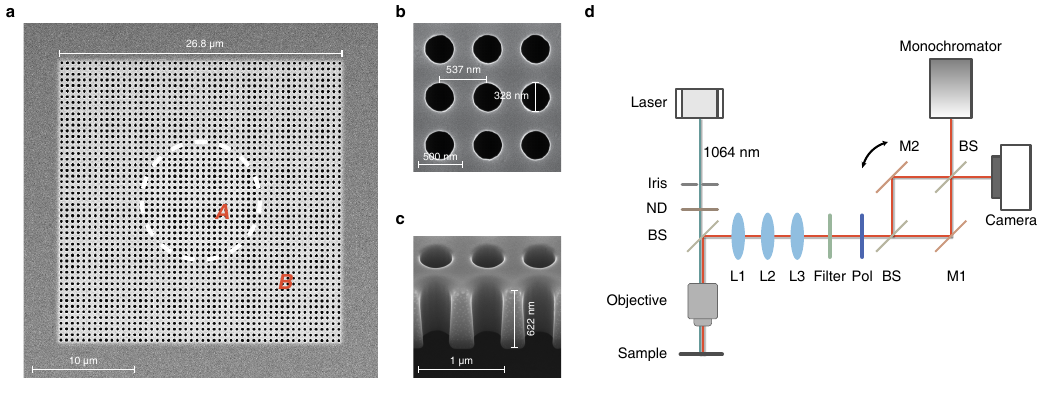}
\caption{\textbf{Fabricated sample and measurement setup}
(a-c)
The scanning electron microscope images of the fabricated sample show the hetero-structures PhC consisting of region A and region B from top and side views. FIB is used to cleave the PhC for a side view of the undercut structure. 
(d)
Schematic of the experimental setup. Iris, iris diaphragm; ND, absorptive neutral density; BS, beam splitter; L, lens; Pol, linear polarizer; M, gold-coated mirror. }
\label{Fig4}
\end{figure}

\clearpage
\begin{figure}[htbp] 
 \centering 
 \includegraphics[width=16cm]{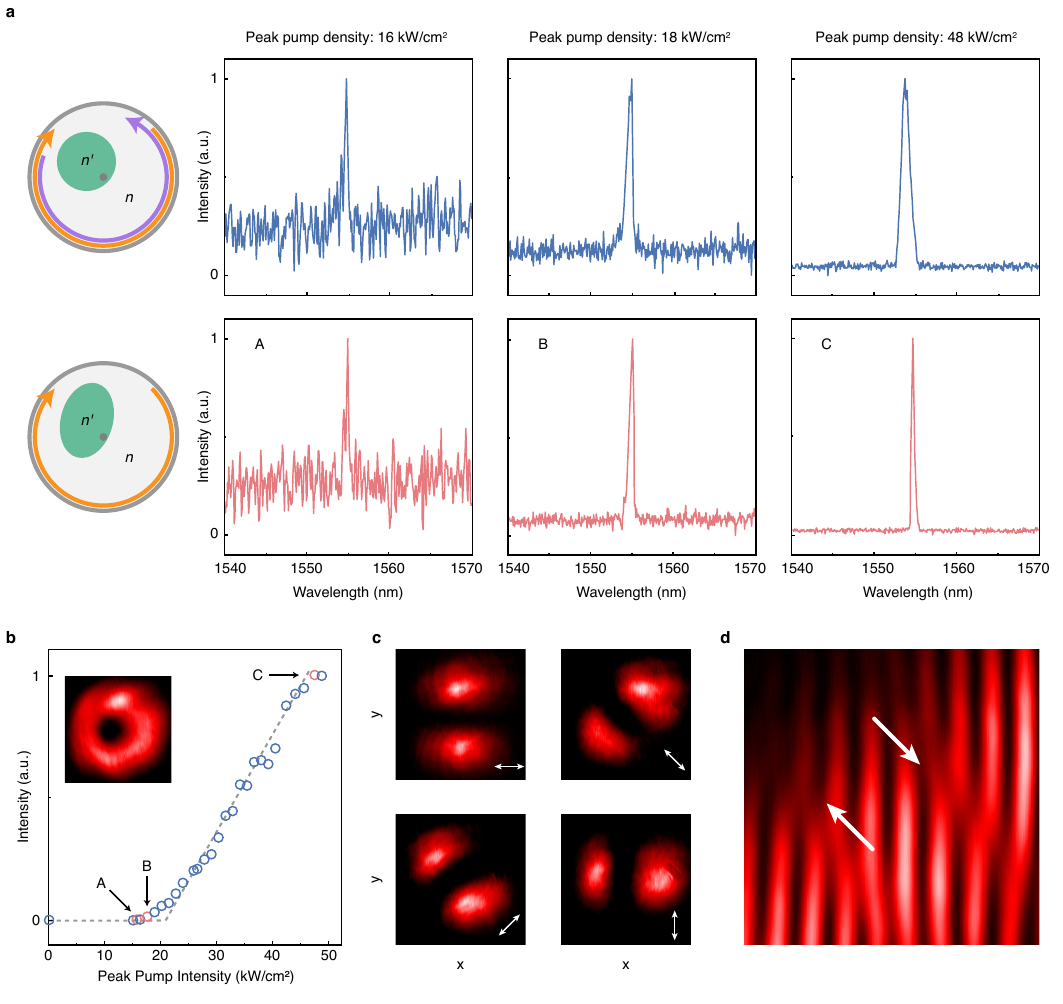}
\end{figure}

\begin{figure}[htbp] 
\caption{\textbf{Characterization of the vortex lasing} 
(a)	The establishing process of lasing oscillation by increasing the pump powers from $16~\mathrm{kW/cm^2}$ to $48~\mathrm{kW/cm^2}$, for symmetric (upper panel) and asymmetric (lower panel) pump conditions, respectively. Three points from A to C are marked, showing the status from spontaneous emission, threshold lasing, to single-mode lasing.
(b)
The power curve of the vortex microlaser is measured, indicating a low threshold of $18~\mathrm{kW/cm^2}$. The inset shows a donut shape of the lasing beam in real space. 
(c)
The polarization-resolved distribution of the vortex beam along $0^\circ$, $45^\circ$,  $90^\circ$, and $135^\circ$, respectively.
(d)
The off-center self-interference pattern of the vortex beam is observed, showing two reversely oriented forks (marked by white arrows) located at the phase singularities as a distinct feature of a phase vortex beam. }
\label{Fig5}
\end{figure}

\end{document}